\documentclass[fleqn,twocolumn,10pt]{wlscirep}
\usepackage{graphicx,float,color}
\usepackage{dblfloatfix,fixltx2e}
\title{Cavity enhanced atomic magnetometry}

\author[1,2,*]{Herbert Crepaz}
\author[1,2]{Li Yuan Ley}
\author[1,2]{Rainer Dumke}
\affil[1]{Centre for Quantum Technologies, National University of Singapore, 3 Science Drive 2, Singapore 117543 }
\affil[2]{Division of Physics and Applied Physics, Nanyang Technological University, 21 Nanyang Link, Singapore 637371}
\affil[*]{hcrepaz@gmail.com}

\begin{abstract}
Atom sensing based on Faraday rotation is an indispensable method for precision measurements, universally suitable for both hot and cold atomic systems. Here we demonstrate an all-optical magnetometer where the optical cell for Faraday rotation spectroscopy is augmented with a low finesse cavity. Unlike in previous experiments, where specifically designed multipass cells had been employed, our scheme allows to use conventional, spherical vapour cells. Spherical shaped cells have the advantage that they can be effectively coated inside with a spin relaxation suppressing layer providing long spin coherence times without addition of a buffer gas. Cavity enhancement shows in an increase in optical polarization rotation and sensitivity compared to single-pass configurations.
\end{abstract}

\begin{document}

\flushbottom
\maketitle
\thispagestyle{empty}

\section*{Introduction}

Among the most sensitive methods to determine the strength of magnetic fields -- surpassing superconducting quantum interference devices\cite{romalis5,fagaly1} -- optical magnetometer have given rise to an increasing range of novel applications\cite{budker1}, including geomagnetism\cite{romalis5}, magnetocardio- and encephalography\cite{prouty1,romalis4,weis2}, quantum nondemolition measurements\cite{romalis3,yabuzaki,mitchell1} and space-borne sensing\cite{burton1}. The development of efficient, high temperature stable antirelaxation coatings\cite{budker3}, spin-exchange free operation\cite{romalis6a} and optical multipass cells\cite{romalis1a, romalis1b} has helped to push sensitivity into the $\mathrm{aT/\sqrt{Hz}}$ regime\cite{romalis6b}. Now there is an increasing demand in developments leading to miniaturized and applicable setups promising portable, battery-operated and highly sensitive devices\cite{kitching1}. Miniaturization while maintaining the interaction strength, and hence sensitivity, leaves one with two options. Either increase atomic density and/or effective optical path length of a probe beam passing the atomic sample. However, a rise in atomic density is accompanied by an increase in spin coherence destroying atom-wall and atom-atom collisions and therefore optimal operating conditions have to be carefully matched to the collisional properties of the atomic species. Multipass cells and cavities are well established methods to increase the photonic contribution to the interaction strength. Until now, only multipass cells had been used due to their feature set of constant path length for each photon, large mode volume and avoidance of optical resonances. Cavities however, may be an attractive choice for miniaturized systems. Chip-scale magnetometer\cite{kitching1} can profit especially as their cell volume and with it sensitivity is more restricted compared to larger table-top systems.
Microcavity fabrication is established technology\cite{vahala1} and less demanding than miniaturizing the more complex geometries of the multipass cells used so far.
Also cavity assemblies can be fit around spherical vapour cells that can be reliably coated with uniform thickness antirelaxation layers, a thing hard to achieve with the edges of cylindrically shaped geometries favoured for multipass cells.

The general measurement principle is based on the modification of optical properties of atoms which are exerted to a magnetic field.
Light that is near resonant with an optical transition creates through optical pumping long-lived atomic ground state polarization moments (orientation, alignment and/or higher order moments), and therefore optical anisotropy in the alkali vapour atoms. In a magnetic field these ground-state spin coherences precess with a field dependent frequency -- the Larmor frequency -- which subsequently can be optically detected. Measured transmission and absorption properties of a probe beam passing the atomic medium are then related to the magnetic field strength the atoms are subjected to.

\begin{figure}[htb]
	\centering
	\includegraphics[width=\linewidth]{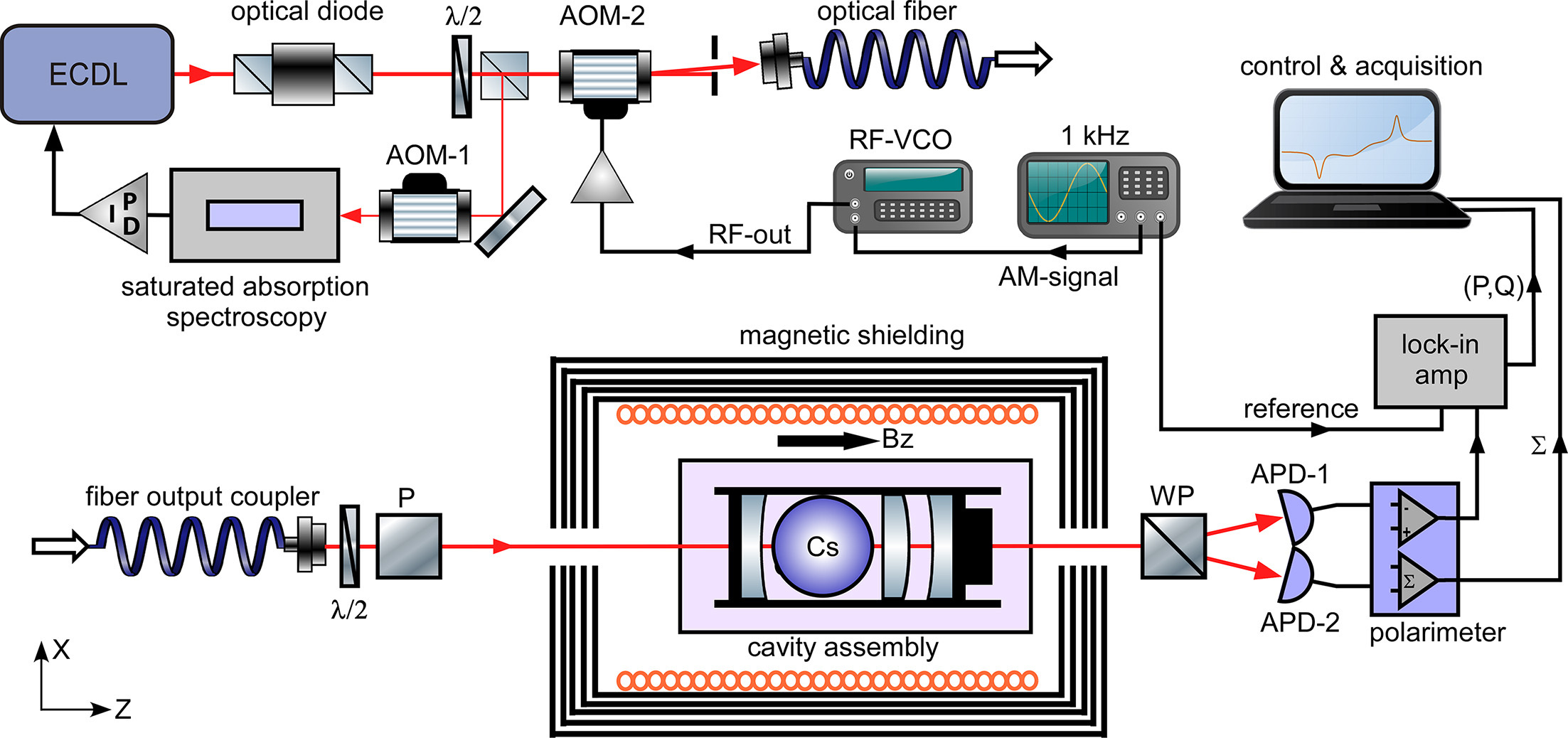}
	\caption{Experimental set-up of the atomic magnetometer. Light from a frequency stabilized external cavity diode laser (ECDL), is intensity modulated by an acousto optic modulator (AOM) and coupled into a single mode polarization maintaining fiber. The polarization of light from the fiber output is adjusted and coupled into the optical cavity assembly which is housed inside a magnetic shield. After the light passes through the cavity the polarization rotation is analysed with a Wollaston prism (WP) and balanced detector. This scheme has been drawn by the author with component symbols from the free ComponentLibrary by Alexander Franzen.}
	\label{fig:setup}
\end{figure}

Here we investigate the potential enhancement of an optical atomic magnetometer with combined pump and probe beam (single-beam configuration) incorporated in a low finesse cavity. In our set-up (see Figure \ref{fig:setup}) the finesse of the cavity was limited by the optical loss within the cavity and the imperfect surface of the spectroscopy cell.
Light is generated by an external cavity diode laser with a typical linewidth below 1 MHz. The laser frequency is stabilized to the $\rm{^{133}Cs}$  D2-transition $\mathrm{6^{2}S_{1/2}\; F=4 \rightarrow 6^{2}P_{3/2}\; F'=4}$ at $\mathrm{\lambda=852.3}$ nm via saturated absorption spectroscopy on an atomic reference cell. Due to the acousto optical modulator (AOM) in the spectroscopy and the AOM before the optical fiber the laser frequency can be shifted within a range of $\pm$100 MHz around the atomic resonance. The second AOM is used to modulate the intensity of the light with a frequency of 1 kHz for lock-in detection of the polarization rotation signal.
After the light has passed through a polarization maintaining single mode fiber the polarization is cleaned by a halve-wave plate and Glan-Thompson polarizer (10$^{5}$ extinction ratio) and then sent into the cavity assembly which is housed inside a 5 layer zero-Gauss chamber. A solenoid placed around the cavity assembly is used to generate a homogeneous magnetic field in the direction of probe beam propagation. The magnetic field inside the chamber can be computer controlled within a range of $\pm$5 mG around zero. After the light has passed through the cavity assembly the polarization rotation is detected via a balanced detection setup comprising a Wollaston Prism (10$^{5}$ extinction ratio) with an axis oriented at 45$^{\circ}$ with respect to the incident light polarization and two avalanche photo detector modules (Thorlabs APD110A/M). The amplified photocurrent signal is then mixed with the intensity modulator reference $\Omega_{m}$ in a two-channel digital lock-in-amplifier and recorded with a computer based high resolution data acquisition system.

\section*{Results}
\subsection*{Intracavity vapour cell.}
\label{cavity_vapor_cell}
To improve the sensitivity of the single-pass configuration we increase the effective interrogation length of the light field with the atomic ensemble. One obvious possibility is to use a multi path geometry. There the beam will be retro-reflected several times at different positions through the cell, and with this the interaction length with the atomic gas will be increased. To go even one step further we followed a different approach and utilized a Fabry-Pérot cavity. In resonance conditions the effective path of the light in the atomic vapour is increased. The effective interrogation length is given by the quality factor of the cavity.

\begin{figure}[htb]
	\includegraphics[width=\linewidth]{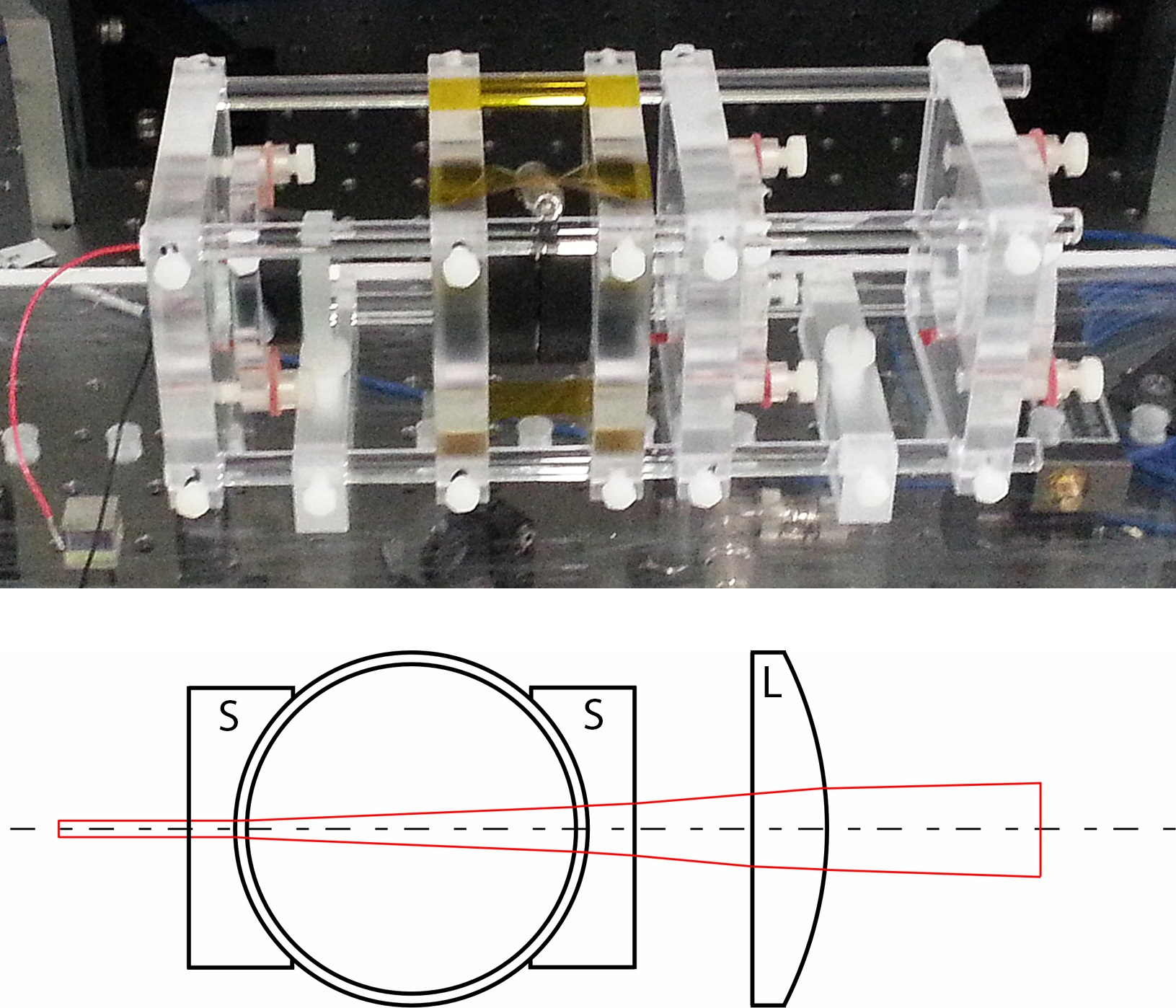}
	\caption{Picture of the cavity assembly (top) and schematic of the optical beam traversing the cavity (bottom). \\The cavity mirrors are mounted in adjustable holders inside a cage system to facilitate easy alignment. The Cs-vapour cell is placed in the center and sandwiched between two AR-coated substrates to minimize reflection losses. An axially adjustable lens, located in front of the exit mirror, allows to mode-match the diverging beam exiting the vapour cell. All parts of the assembly are made of nonmagnetic polymers or glass.}
	\label{fig:cavity}
\end{figure}

We have designed a Fabry-Pérot cavity encapsulating the anti relaxation coated Cs cell\cite{weis1}, as indicated in Figure \ref{fig:cavity}. The cell is at ambient temperature ($ \sim 25^{\circ} C$) corresponding to a density of $\mathrm{n=6\cdot 10^{10}\, atoms/cm^{3}}$. Besides the atomic vapour cell additional optical elements have to be introduced into the cavity assembly (see Methods section\ref{ncs}). 
The main loss mechanism are still reflections from the inner surface of the anti relaxation coated vapour cell. This loss determines in this set-up the finesse of the enhancement cavity.

With the above described configuration we can achieve a measured finesse of the cavity of 12.7 as we expected from the design. The free spectral range is 954 MHz and the line width is 75 MHz\ref{cavity_calibration}. This will also increase the effective path length by a factor of $\mathrm{\sim 10}$. However at the same time the light intensity is increased inside the cavity which would lead to a power broadening of the atomic signal if we would use the same intensities used in the single-pass magnetometer. To counter the decrease in detection signal by lowering the probe light power (0.8 $\mathrm{\mu W}$) avalanche photodiodes were implemented.

\subsection*{Cavity augmented AM-NMOR signal.}
\label{NMOR}

To measure magnetic field induced optical rotation we employ Bell-Bloom\cite{bloom} type synchronous optical pumping and probing with amplitude modulated light (AMOR). This creates atomic alignment, where atoms are pumped into a stationary spin-state when viewed in a frame rotating with twice the Larmor frequency $\Omega_{L}$. Linear polarized light pumps multilevel atoms into states with even polarization moments. The lowest order even polarization moment is termed alignment. Due to the twofold symmetry resonance occurs when $\Omega_{m}=\pm 2 \Omega_L$. For circular polarized pump light orientation is created which is resonant when $\Omega_{m}=\pm \Omega_L$.
When the modulation frequency equals two times the Larmor frequency $\mathrm{\Omega_{m}=\pm 2 \Omega_L=\pm 2 g \mu_{B} B/\hbar}$, where g denotes the Landé factor and $\mu_{B}$ the Bohr magneton, a resonance in the transmitted light is observed. Frequency modulation, which can be employed alternatively to amplitude modulation in standard NMOR-setups, is not suitable in this case because the required modulation depth (300 MHz-1 GHz) would be much larger than the cavity linewidth (75 MHz).

\begin{figure}[htb]
	\includegraphics[width=\linewidth]{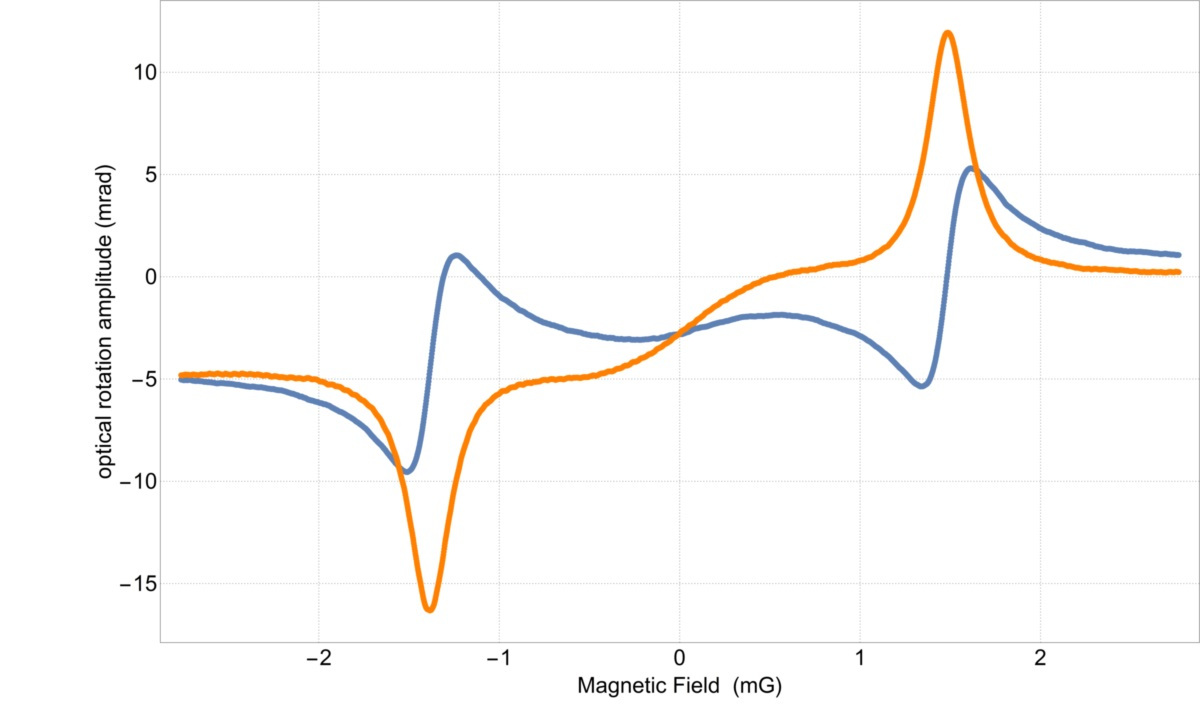}
	\caption{Nonlinear optical rotation signal as a function of longitudinal magnetic field.
		In-phase (blue) and out-of-phase (red) component of the NMOR signal. The symmetric line-shapes appears due to optical pumping of atoms synchronized with the modulation frequency $\Omega_{m}$. The central resonance around B$\approx$0, the typical nonlinear Faraday signal, is strongly suppressed compared to conventional single beam AMOR signals (see Methods section).}
	\label{fig:NMOR-signal}
\end{figure}

Figure \ref{fig:NMOR-signal} presents a measurement of optical rotation dependent on the longitudinal magnetic field demodulated at the first harmonic of $\Omega_{m}$. For this measurement a 0.9 mm diameter probe beam with a time averaged intensity of 1.3 $\mathrm{\mu W/mm^{2}}$, tuned +80 MHz off resonance of the $\rm{6^{2}S_{1/2}\; F=4 \rightarrow 6^{2}P_{3/2}\; F'=5}$ transition and sinusoidal amplitude modulation with 100 $\%$ modulation index was used. Sinusoidal modulation is a good compromise between the competing duty-cycle requirements for optimal optical pumping and probing\cite{gawlik2}. To reduce current noise from the power supply driving the solenoid coil data acquisition was phase-locked to the 50 Hz AC mains powerline.  
 
\subsection*{Noise measurements.}
\label{Noise}

For the noise spectra measurements the magnetometer signal was recorded on the centre of the dispersive slope of the $-1^{st}$ order in-phase quadrature component of the lock-in signal located around -1.4 mG (see blue graph in Figure \ref{fig:NMOR-signal}). The lock-in amplifier (Femto LIA-MVD-200H) output signal was digitized and subsequently Fourier transformed to yield the voltage power spectrum. Conversion of the voltage noise to magnetic field noise is determined by the slope of the dispersive optical rotation signal as a function of the B-field. All optical elements were enclosed in boxes to reduce low frequency ($<$ 10 Hz) drifts of the optical rotation signal caused by beam steering from air convection.
From a fit to the dispersive signal the slope factor calculates to $\delta B / \delta U = 20.3$ nT/V and the optical rotation calibration yields $ \delta \phi /\delta B  = 42.16$ mrad/mG.

\begin{figure}[htb]
	\includegraphics[width=\linewidth]{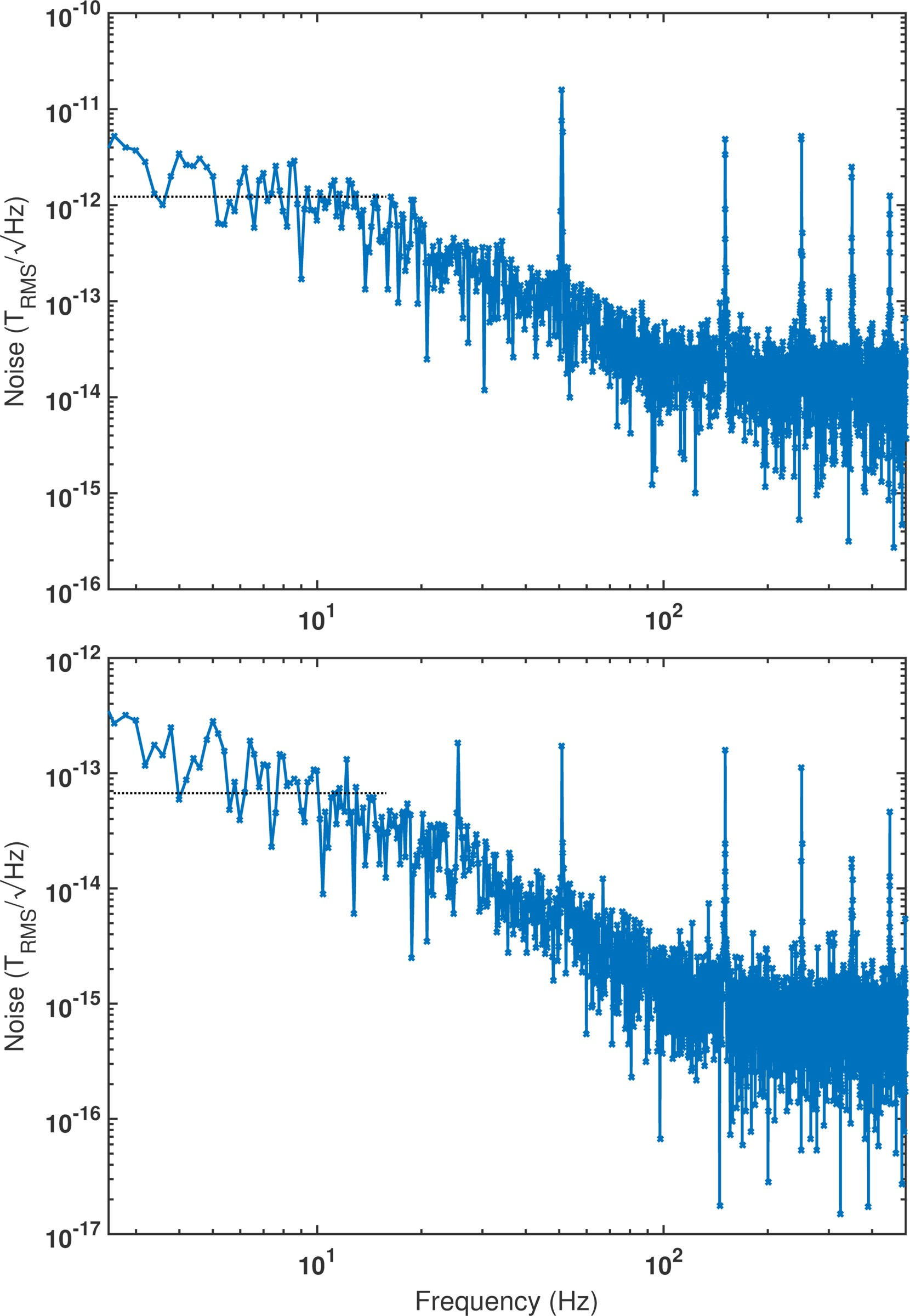}
	\caption{Root power spectral density of the magnetometer signal without cavity (top) and with cavity (bottom). The graph shows the demodulated magnetometer response at twice the Larmor frequency of 1 kHz. AC-noise sidebands appear at multiples of 50 Hz due to imperfect shielding and residual current noise from the power supply driving the solenoid coil. The raw data were recorded with a low noise 18-bit ADC with 1000 Hz sample rate and then Fourier transformed. Note that for the arrangement without cavity the photo detector system and probe beam parameters are different compared to the cavity set-up. Dashed lines indicate rms noise level of data points residing within a 1 Hz wide frequency window centred at 10 Hz.}
	\label{fig:noise_spectrum_cavity}
\end{figure}

A comparison of the magnetic noise spectrum in a single pass set-up with (bottom graph) and without cavity enhancement (top graph) is depicted in Fig.\ref{fig:noise_spectrum_cavity}. Note that the operating conditions eg. beam size and beam power and the detection system were optimized for each set-up individually and are therefore different. Comparison of the low-frequency signal levels at 10 Hz of $\mathrm{1.2\cdot 10^{-12}\; T/\surd Hz}$ (dashed line, top) for the single-pass setup and $\mathrm{6.9\cdot 10^{-14}\; T/\surd Hz}$ (dashed line, bottom) for the cavity enhanced one shows that the sensitivity of cavity enhanced interrogation of the atoms is increased by a factor of $\mathrm{\sim 17}$ with respect to the conventional configuration without cavity. The signal contributions at 10 Hz used for comparison had been chosen to lie within the 17 Hz (-3 dB) filter bandwidth of the low-pass filter of the lock-in amplifier. Signal levels had been estimated from the rms values of 10 equally spaced data points located within a frequency window of $\pm 0.5$ Hz centred at 10 Hz.
Our measured increase in sensitivity is slightly higher than the cavity finesse of $F=12.7$. For a purely dispersive, nonabsorbing media cavity enhancement of the Faraday effect has been predicted to be proportional to the cavity finesse\cite{ling1}. Thus one might be tempted to increase the finesse significantly to gain proportionally enhanced sensitivity. However, for resonantly absorbing media, as it is the case for linear and nonlinear Faraday rotation, higher intracavity power may lead to increased induced intracavity losses and power broadening of the atomic transitions involved\cite{sycz1}. Keeping intracavity losses and absorption low while avoiding power broadening is crucial. These requirements become increasingly challenging for higher finesse cavities and will set limits to the effectivity of cavity enhancement at high finesse values.  
 

\begin{figure}[htb]
	\includegraphics[width=\linewidth]{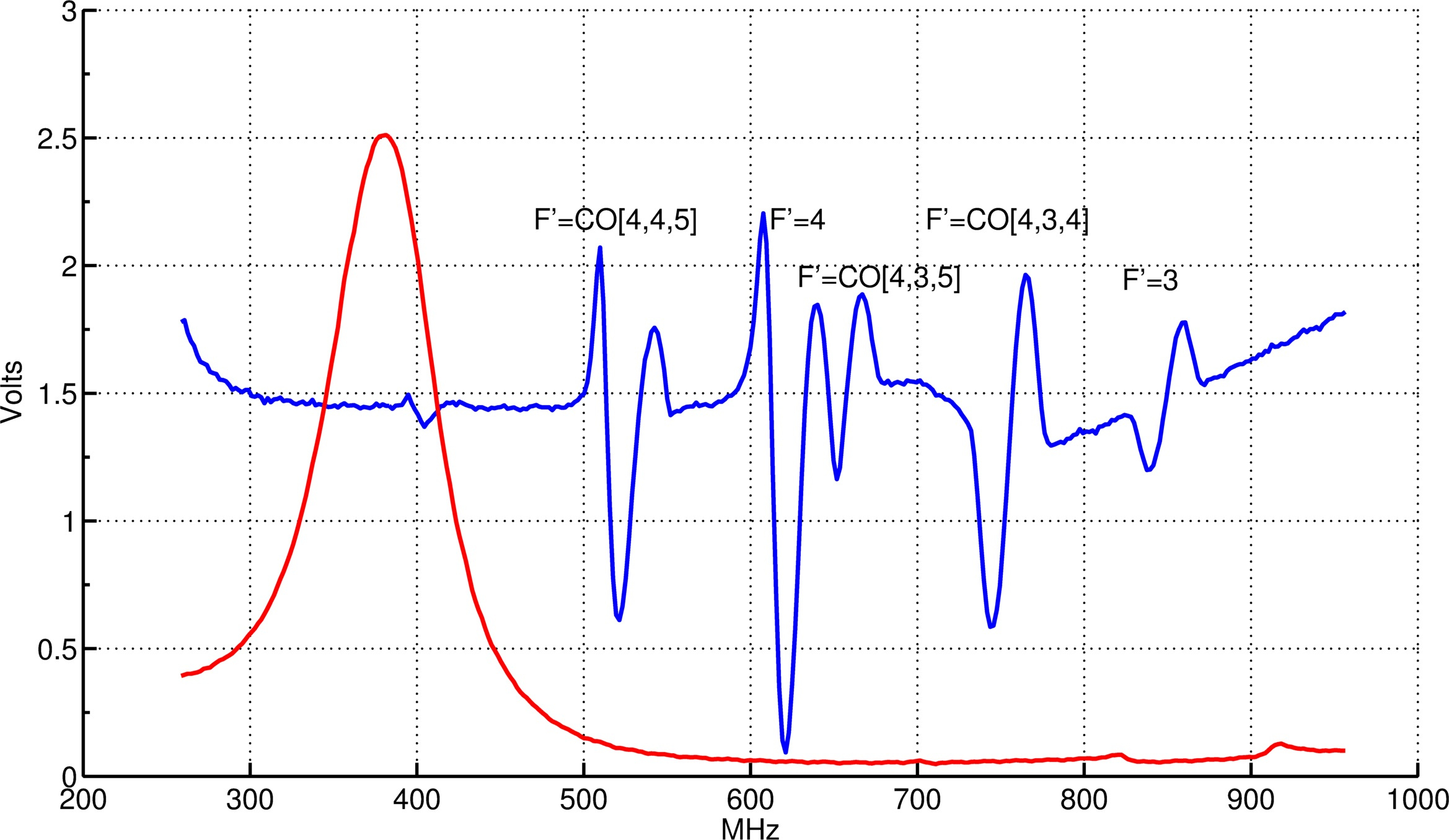}
	\caption{Measurement of the cavity transmission linewidth.\\
		Frequency scan of the cavity assembly transmission signal (red) and error-signal (blue) from a $^{133}$Cs D2-line saturated absorption spectroscopy. The spacings between the zero crossings of the $\mathrm{6^{2}S_{1/2},\,F=4 \rightarrow 6^{2}P_{3/2},\,F^{\prime}=3,4,5 }$ spectroscopy signal serve as absolute frequency marker for calibration. From this the measured cavity linewidth calculates to $\mathrm{\delta \nu  = 2\pi*75}$ MHz (FWHM). The signal amplitude in Volt is proportional to the photocurrent of the polarimeter.}
	\label{fig:cavity-transmission}
\end{figure}

\section*{Discussion}

In this work we presented a highly sensitive, atomic magnetometer based on a cavity augmented Cs vapour cell. Here, enhanced interaction leads to an increased optical rotation and sensitivity compared to an otherwise equivalent single pass setup. Although cavity enhancement to increase the optical path-length and therefore atom-light interaction strength is well established for precision experiments\cite{hinds1}, it has not been used so far in optical magnetometry where multipass cells had been favoured instead\cite{romalis1a,romalis1b}. The absence of optical resonances, a large mode volume and constant path length for each photon makes them attractive. For certain applications however, cavities may be preferable. Compared to the complex geometry of multipass cells used for optical magnetometer in recent work, a low finesse cavity can be arranged around antirelaxation coated standard spherical vapour cells thereby taking advantage of the long spin coherence times coated cells offer. Also, cavities lend themselves better to miniaturization. Fabrication of microcavities is established technology and indispensable for areas as diverse as optical telecommunication and cavity QED\cite{vahala1,hinds1}. In chip-scale atomic magnetometer\cite{kitching1}, where the probe volume and hence sensitivity is limited, coupling to microfabricated cavities may be especially beneficial.

\section*{Methods}
\label{methods}
{\small
\subsection*{Nonmagnetic cavity setup.}
\label{ncs}
Two anti reflection coated plano-concave substrates house the Cs-vapour cell. The concave surface is contacted to the cell with refractive index matching gel. Using the vapour cell directly would lead to a decrease in the Quality factor of the cavity, since approx. 4\% of light on each surface would be reflected and lost from the cavity mode per round trip. Furthermore the inhomogeneous outer surface of the vapour cell would degrade the mode quickly. This is counteracted by placing these two substrates with optical quality surfaces.
The atomic cell itself with the attached substrates, still acts as a type of ball lens. To compensate this effect an additional anti reflection coated plano-convex lens is placed inside the cavity. This arrangement of optical components and cavity mirrors resembles an effective hemispherical resonator.
Since the cavity will be part of the highly sensitive magnetometer it has to be assembled from selected non magnetic materials. Their magnetic properties were carefully experimentally verified to not interfere with the magnetometer operation. The housing for the cavity and holders for the plano-convex substrates where 3D printed using thermo plastic. The main body of the mirror and lens holders were made of acrylic (PMMA - polymethyl methacrylate). To adjust the angle of the mirrors and lenses a design utilizing Nylon screws and flexible rubber strings was implemented. 
A piezo attached to one of the cavity mirrors allows active stabilization of the cavity length prior to a measurement. During a measurement, which lasts about 5-10 seconds, the low finesse cavity is sufficiently stable to be operated in passive mode without active stabilization. For cavities with a higher finesse where drift stability is crucially important active stabilization could, in principle, be periodically applied during the off-cycles of the amplitude modulated pump-probe beam.   

\subsection*{Calibration of the cavity transmission signal.}
\label{cavity_calibration}
Figure \ref{fig:cavity-transmission} shows the transmission signal of a probe beam passing the cavity assembly. Frequency adjustment of a probe beam periodically changes the transmitted light intensity according to the Airy function dependence $I_{T}=(\frac{I_{0}}{1+F \sin(\delta/2)}).$ Here $F$ denotes the coefficient of Finesse and $\delta$  the optical path length. To get an absolute frequency reference a saturated absorption spectrum is recorded during probe beam detuning. The known separations between optical transitions in the spectrum are then used to calibrate the cavity free spectral range  $FSR = 2\pi*954$ MHz and linewidth  $\delta \nu  = 2\pi*75$ MHz (FWHM) leading to a finesse of $F=12.7$.

\subsection*{Shape of the AMOR signal.}
\label{AMOR_shape}
  The line shape of the demodulated optical rotation signals seen in Figure \ref{fig:NMOR-signal} is caused by the phase shift of the forward scattered light field induced by atomic alignment of the ground state and can be well described by a dispersive Lorentzian profile\cite{Malakyan}.\\ 
 In the regime around $B_{z}\sim 0$, where $\Omega_{L}<<\Omega_{m}$ and $\Omega_{L}<\Gamma_{relax}$, with $\Gamma_{relax}$ denoting the relaxation rate of ground state coherences, alignment relaxes before a full period of optical rotation occurs. In our setup relaxation is dominated by depolarization due to atom-wall and spin-exchange collisions. The measured relaxation rate for our paraffin coated cell at room temperature is $\Gamma_{relax}=2\pi*4.4$ Hz (data from Prof. A. Weis, Université de Fribourg).
 When $\Omega_{L}\sim\Gamma_{relax}$, average atomic alignment has its axis rotated with respect to the incident light polarization by an angle $\phi<\pi/4$ and therefore causes optical rotation and the appearance of a resonance\cite{budker1}. Here, in contrast to conventional single-beam AMOR signals, the central resonance around $B_{z}\sim 0$ is strongly suppressed.  We attribute the disappearance of this resonance in our set-up to the unique conditions inside a cavity enhanced cell. Atoms moving through the cavity field experience optical pumping at different times depending on their position within the cavity modes leading to a dephasing and averaging out of the collective magneto-optical rotation signal. 
 The increase in the vertical offset of the $1^{st}$-order resonances at increasing B-fields in Figure \ref{fig:NMOR-signal} is caused by the transit time limited Faraday effect\cite{gawlik2}. There atoms are pumped, precess and are interrogated during a single pass through the cavity field. The measured background rotation factor of $ \delta \phi /\delta B  = 0.588$ mrad/mG from the transit effect resonance has a width of $\sim$71 mG limited by the transit rate $\Gamma_{t}=2\pi*25$ kHz.

}

\section*{Acknowledgements}
We thank the group of Antoine Weis, Université de Fribourg for the production of two paraffin coated Cs vapor cells.\\
This work was financially supported by DSO National Laboratories, Singapore and Centre for Quantum Technologies under project agreement DSOCL12111.

\section*{Author contributions statement}
 H.C. and R.D. conceived the experiment and drafted the manuscript. L.Y.L. constructed the apparatus and performed the measurements. H.C. analysed the data and supervised design and construction. R.D. supervised the whole project.

\section*{Additional information}

\subsubsection*{Competing financial interests} 
The authors declare no competing financial interests.

\subsubsection*{Corresponding author}
Correspondence to: Herbert Crepaz or Rainer Dumke
 
\newpage

\end{document}